\documentclass[aps,prl,twocolumn,showpacs,superscriptaddress]{revtex4-1}

\usepackage[pdftex]{graphicx} 
\usepackage{natbib} 

\newcommand{\bv}[1]{\mathbf{#1}}
\newcommand{\bvv}[1]{{\bf#1}}

\newcommand{\srI}{Sr$_{2}$RuO$_{4}$}
\newcommand{\srII}{Sr$_{3}$Ru$_{2}$O$_{7}$}

\newcommand{\odyz}{$d_{{yz}}$}
\newcommand{\odxz}{$d_{{xz}}$}

\newcommand{\zfl}{$Z_{FL}^{-1}$}

\begin{document}

\title{Formation and consequences of heavy $d$--electron quasiparticles in \srII}

\author{M.P. Allan}
\affiliation{SUPA, School of Physics and Astronomy, University of St Andrews, St Andrews, Fife KY16 9SS, United Kingdom}
\affiliation{LASSP, Department of Physics, Cornell University, Ithaca NY 14850, USA}

\author{A. Tamai}
\affiliation{SUPA, School of Physics and Astronomy, University of St Andrews, St Andrews, Fife KY16 9SS, United Kingdom}

\author{E. Rozbicki}
\affiliation{SUPA, School of Physics and Astronomy, University of St Andrews, St Andrews, Fife KY16 9SS, United Kingdom}

\author{M.H. Fischer}
\affiliation{LASSP, Department of Physics, Cornell University, Ithaca NY 14850, USA}

\author{J. Voss}
\affiliation{School of Applied and Engineering Physics, Cornell University, Ithaca NY 14850, USA}

\author{P.D.C. King}
\affiliation{SUPA, School of Physics and Astronomy, University of St Andrews, St Andrews, Fife KY16 9SS, United Kingdom}

\author{W. Meevasana}
\affiliation{SUPA, School of Physics and Astronomy, University of St Andrews, St Andrews, Fife KY16 9SS, United Kingdom}
\affiliation{School of Physics, Suranaree University of Technology, Nakhon Ratchasima, 30000, Thailand}

\author{S. Thirupathaiah}
\affiliation{Helmholtz-Zentrum Berlin, Elektronenspeicherring BESSY II, D-12489 Berlin, Germany}

\author{E. Rienks}
\affiliation{Helmholtz-Zentrum Berlin, Elektronenspeicherring BESSY II, D-12489 Berlin, Germany}

\author{J. Fink}
\affiliation{Helmholtz-Zentrum Berlin, Elektronenspeicherring BESSY II, D-12489 Berlin, Germany}
\affiliation{IFW Dresden, P.O. Box 270116, D-01171 Dresden, Germany}

\author{A. Tennant}
\affiliation{Helmholtz-Zentrum Berlin, Hahn-Meitner-Platz 1, D-14109 Berlin, Germany}

\author{R.S. Perry}
\affiliation{SUPA, School of Physics and Astronomy, University of St Andrews, St Andrews, Fife KY16 9SS, United Kingdom}

\author{J.F. Mercure}
\affiliation{SUPA, School of Physics and Astronomy, University of St Andrews, St Andrews, Fife KY16 9SS, United Kingdom}

\author{M.A. Wang}
\affiliation{LASSP, Department of Physics, Cornell University, Ithaca NY 14850, USA}

\author{C.J. Fennie}
\affiliation{School of Applied and Engineering Physics, Cornell University, Ithaca NY 14850, USA}

\author{E.-A. Kim}
\affiliation{LASSP, Department of Physics, Cornell University, Ithaca NY 14850, USA}

\author{M.J. Lawler}
\affiliation{Department of Physics, Binghamton University, Binghamton NY, 13902, USA}
\affiliation{LASSP, Department of Physics, Cornell University, Ithaca NY 14850, USA}

\author{K.M. Shen}
\affiliation{LASSP, Department of Physics, Cornell University, Ithaca NY 14850, USA}

\author{A.P. Mackenzie}
\affiliation{SUPA, School of Physics and Astronomy, University of St Andrews, St Andrews, Fife KY16 9SS, United Kingdom}

\author{Z.-X. Shen}
\affiliation{Departments of Applied Physics, Physics, and Stanford Synchrotron Radiation Laboratory, Stanford University, Stanford, California 94305, USA}

\author{F. Baumberger}
\affiliation{SUPA, School of Physics and Astronomy, University of St Andrews, St Andrews, Fife KY16 9SS, United Kingdom}


\begin{abstract}
We report angle-resolved photoelectron spectroscopy measurements of the quantum critical metal \srII{} revealing itinerant Ru 4$d$-states confined over large parts of the Brillouin zone to an energy range of $\lesssim 6$~meV, nearly three orders of magnitude lower than the bare band width. We show that this energy scale agrees quantitatively with a characteristic thermodynamic energy scale associated with quantum criticality and illustrate how it arises from the hybridization of light and strongly renormalized, heavy quasiparticle bands. For the largest Fermi surface sheet we find a marked \bvv{k}-dependence of the renormalization and show that it correlates with the Ru~4$d$ - O~2$p$ hybridization.
\end{abstract}

\maketitle

The bilayer ruthenate \srII{} exhibits intriguing thermodynamic and transport properties -- including itinerant metamagnetism and electronic nematicity \cite{per01,gri01,gri04,bor07,far08,ros09,ros11} -- that are reminiscent of $f$-electron quantum critical heavy-fermion systems and have generated significant theoretical interest \cite{bin04,kee05,yam07,rag09,ber09,lee09a,fis10,pue10,fra11,pue12}. 
These properties are remarkably different from the single layer strontium ruthenate \srI\ and cannot be understood in an independent electron picture. Within band structure theory in the local density approximation (LDA), \srI\ and \srII\ have a wide bare conduction band, formed by relatively extended Ru~4$d$ states hybridizing with O~2$p$ electrons, and a similar density of states at the Fermi level. Despite the moderate Coulomb repulsion in the Ru~4$d$ shell, a sizable correlation driven enhancement of the Sommerfeld coefficient of $\gamma/\gamma_{\rm{LDA}}\approx 4$ is observed in \srI\ which was recently argued to be  a consequence of Hund's rule coupling reducing the coherence scale away from half filling \cite{mra11,med11}. The even larger electronic specific heat of \srII\ with $\gamma/\gamma_{\rm{LDA}}\approx 10$ is indicative of enhanced electronic correlations in the bilayer system.

Recent transport and entropy data give a detailed picture of quantum criticality and the formation of an electron nematic phase in \srII\ \cite{bor07,ros09,mer10,ros11}. Particularly relevant in the context of this paper is the observation of a maximum in the electronic specific heat $C_{\rm{el}}(T)/T$ near 8K with its position continuously suppressed with increasing field, terminating in a logarithmic divergence at the putative quantum critical end point \cite{ros11}. Taken together with the absence of a mass divergence in several Fermi surface sheets \cite{mer10}, this provides strong evidence that criticality is driven by the suppression of a single energy scale, which is at least two orders of magnitude smaller than the bare band width, and present in a subset of bands only.
However, despite recent progress in characterizing the low-energy electronic structure of \srII\ \cite{iwa07, tam08, lee09}, the microscopic origin of this energy scale remained elusive. 

\begin{figure*}[ht!]
\includegraphics[width=18 cm]{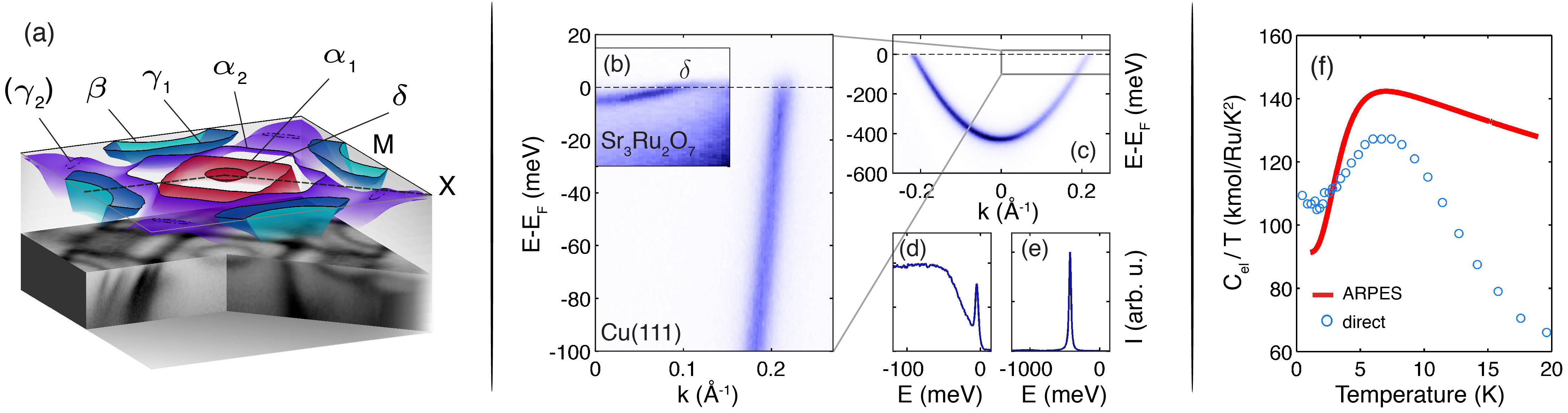}
\caption{(a) Parametrization of the low-energy quasiparticle band structure of \srII{} derived from experimentally determined initial state energies. (b-c) False color plots 
(dark colors correspond to high photocurrent throughout this paper) from the L-gap surface state on Cu(111) and the $\delta$-pocket on \srII, illustrating the marked effect of correlations on the quasiparticle dispersion in \srII. (d,e) Spectra at the $\Gamma$ point of \srII\ and Cu(111), respectively. (f) Comparison of the zero-field electronic specific heat from Ref.~\cite{ros11} with a calculation of $C_{el}(T)/T=\frac{{1}}{{T}}\frac{{\partial}}{\partial{T}}\int \varepsilon g(\varepsilon)f(\varepsilon,T)d\varepsilon$ based on ARPES data.
The density of states $g(\varepsilon)$ is computed numerically from the parametrization of the experimental low-energy band structure shown in (a).
We assumed a two-fold degeneracy of the bands around the X-point, as indicated in Fig.~2(a) and a constant $g(\varepsilon)$ above the Fermi level.}
\end{figure*}

Here, we report detailed angle-resolved photoelectron spectroscopy (ARPES) measurements of \srII\ revealing flat Ru 4$d$ bands that define an energy scale consistent with thermodynamic measurements.  We show how this energy scale arises microscopically from the hybridization of strongly renormalized bands with dispersive states.  
The resulting Fermi liquid has a strongly sheet \textit{and} momentum dependent renormalization of the Fermi velocity, reaching values more typically encountered in $f$-electron Kondo systems.

For the experiments presented here, we used crystals grown by the floating-zone method as described in Ref.~\cite{per04a} with residual resistivities as low as 0.4~$\mu\Omega$cm. Our photoemission experiments  were performed  using 16~to 57~eV photons  from SSRL's beamline V-4, BESSY-II's $1^3$ beamline and the SIS beamline at SLS, as well as He~I$\alpha$ radiation from monochromatized discharge lamps. The measurements were taken at temperatures around 8~K (Figs.~1, 2(a,e), 3) and 1.1~K (Figs.~2(a,b,c)) and energy and angular resolutions of 3.5 to 5~meV and $\approx 0.3^{\circ}$, respectively.  Density functional calculations within the local density approximation (LDA) were performed using the all-electron code Wien2k including spin-orbit coupling and the pseudo-potential code Quantum Espresso~\cite{bla2k,gia09ea}.

\begin{figure}[!hbt]
\includegraphics[width=8.7cm]{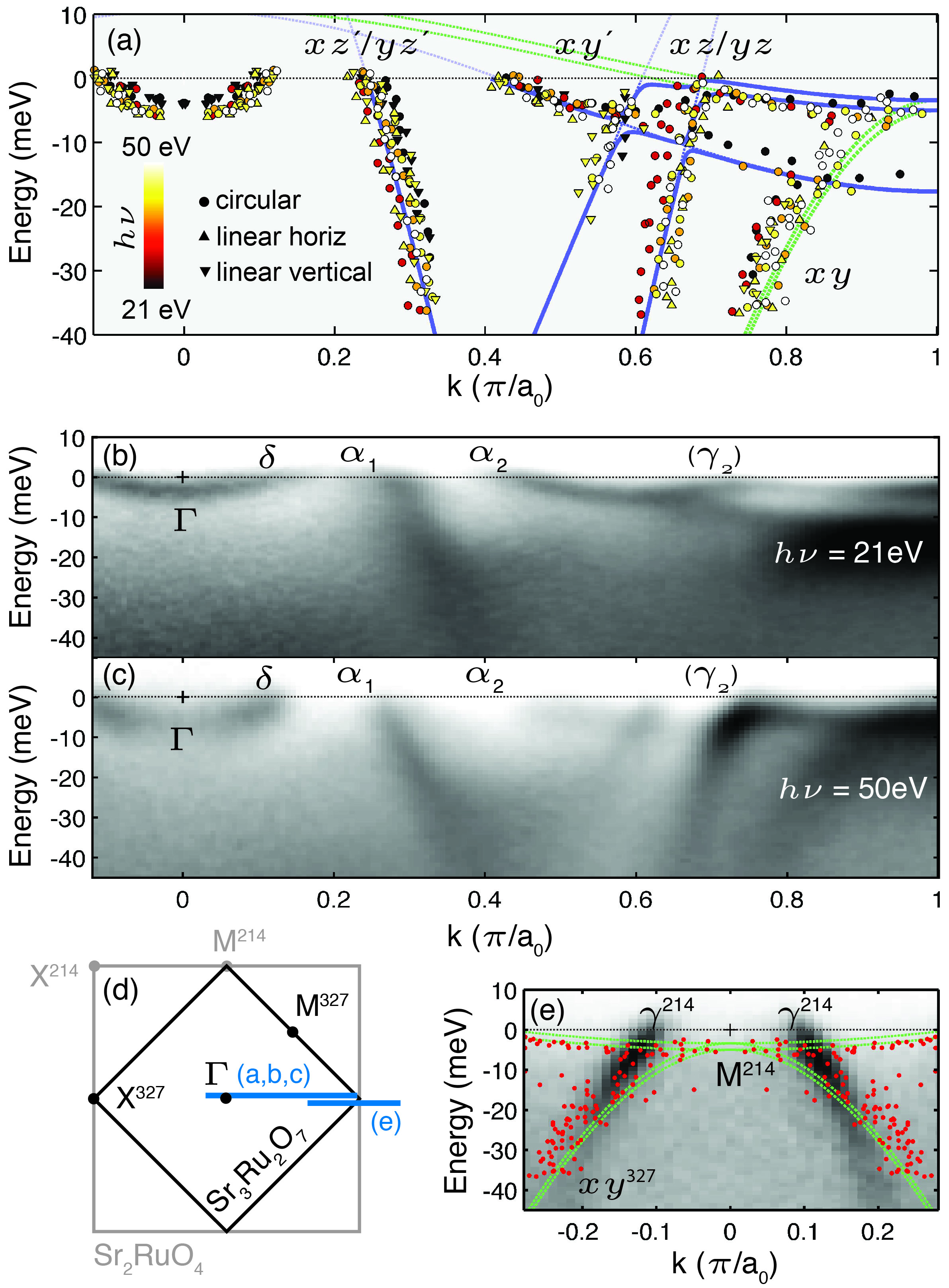}
\caption{(a) Band dispersion along $\Gamma$X extracted from ARPES measurements with different photon energies and polarizations. 
The parallel momentum is given in units of $\pi/a$, where $a$ is the lattice constant in the undistorted tetragonal structure of \srI.
8 cosine dispersions track the fundamental $xz/yz$ and $xy$ bands and their backfolded copies ($xz/yz'$, $xy'$). The low-energy contours arising from the hybridization of these bands is indicated by a thick blue line.
(b,c) Representative cuts measured with $h\nu=21.2$~eV and 50~eV photons, respectively. 
(d) Relation of the Brillouin zones of \srI\ and \srII.
(e) Comparison between the $xy$ band in \srI\ (false color plot) and \srII\ (data points and green lines from panel (a)).}
\end{figure}

The low-energy electronic structure of \srII\ is summarized in Fig.~1(a) where we show the experimental Fermi surface data from our previous work \cite{tam08} with a new parametrization of the low-energy quasiparticle dispersion derived from $\approx 10^3$ experimentally determined initial state energies. While the band topography resembles the LDA band structure \cite{sin01, tam08}, the experimental quasiparticle velocities are markedly lower than the calculated bare velocities. 
This behavior is well known for correlated Fermi liquids and can be characterized by a renormalization constant (Fermi liquid residue) $Z_{FL}^{-1} \approx v_{\rm{LDA}}/v_{\rm{exp}}$, where $v_{\rm{LDA}}$ and $v_{\rm{exp}}$ are the group velocities of bare bands as calculated within LDA and the measured quasiparticle bands at the Fermi level, respectively. The magnitude of \zfl\ in \srII\ is highly unusual even for transition metal oxides, and contrasts strongly with a weakly interacting electron gas. We illustrate this  in Fig.~1(b-e), where we compare the parabolically dispersing $\delta$-pocket of \srII{} with the weakly interacting electron gas found at the Cu(111) surface \cite{gar75, rei01}. Both of these states have similar bare band masses of $\approx 0.4$~m$_e$. Yet, while the quasiparticle dispersion in Cu closely follows the bare band, the band width of the $\delta$-pocket in \srII\ is reduced to $\approx 5$~meV, corresponding to a renormalization constant of \zfl~$\approx25$.
At the same time the spectral weight of the coherent quasiparticle peak is reduced and pushed to higher energy, resulting in the characteristic `peak-dip-hump' line shape of strongly interacting systems (Fig.~1(d)).

Equally strong band renormalization, combined with marked multi-band effects, is observed over a large $k$-space volume spanned by the hybridized $\alpha_2 - \gamma_2$ sheet. For these bands of mixed $xz/yz$, $xy$ orbital character, the low energy scale is much more consequential, since the larger area in $k$-space leads to a higher associated density of states. In Fig.~2 we show the quasiparticle dispersion along $\Gamma$X extracted from several measurements with different photon energies and polarizations. 
Attempting to describe the experimental dispersion with a minimal model, we approximate the peak positions by 8 cosine bands tracking the dispersions of the fundamental bilayer split $xy$ and $xz/yz$ orbitals and their back-folded copies $xy'$, $xz/yz'$. [See Fig.~2(d) for a sketch of the reduced Brillouin zone of \srII. The bilayer splitting in the $xy$ sheet is not resolved experimentally and is for illustrative purposes only.]
Intriguingly, these cosine bands have very different band widths with Fermi velocities varying by more than an order of magnitude. In the presence of spin-orbit coupling \cite{liu08,hav08} this situation naturally leads to the hybridization of very itinerant light bands and heavy states near the Fermi level.
The resulting low-energy contour (Fig.~2(a)) remains confined to an energy range of $\lesssim 6$~meV over an extended part of the Brillouin zone and has a complex shape with multiple saddle-point and band-edge singularities, mostly close to the X point. This includes a band maximum at $-1\pm1$meV that might lead to the putative $\gamma_2$ pocket, as discussed in our previous work~\cite{tam08}.

Several authors proposed that many (although not all) properties of quantum critical materials as they are seen in \srII\ and some heavy fermion systems can be explained assuming a narrow peak in the density of states close to or locked to the chemical potential \cite{far08,ros11,dao06,pfa12}. Moreover, Rost \emph{et al.}~\cite{ros11} recently showed experimentally that the quantum critical endpoint in \srII\ appears to be rooted in the suppression of a single energy scale, identified as the peak position of a maximum in the zero-field specific heat $C_{\rm{el}}(T)/T$ around 8~K. 
The heavy bands observed by ARPES naturally lead to a $C_{\rm{el}}(T)/T$ curve similar to the one reported in Ref.~\cite{ros11}. Indeed, a calculation of $C_{\rm{el}}(T)/T=\frac{{1}}{{T}}\frac{{\partial}}{\partial{T}}\int \varepsilon g(\varepsilon)f(\varepsilon, T)d\varepsilon$ using the quasiparticle density of states $g(\varepsilon)$ obtained from a histogram of the parametrized low-energy band dispersion shown in Fig.~1(a) reproduces the hump in $C_{\rm{el}}(T)/T$ at the correct temperature (Fig.~1(f)), as a direct consequence of the low energy contours discussed above. 

This strongly suggests that the energy scale defined by the hybridized $\alpha_2$ - $\gamma_2$ band, which dominates $g(\varepsilon)$, is intimately involved in quantum criticality and the formation of an electron nematic state in high field. 
The complex shape and orbital character of this sheet, evolving from the out-of-plane $\alpha_2$ pocket to $xy$-dominated states near X$^{\rm{327}}$ (as defined in Fig.~2(d)), highlights the need for realistic models of nematicity in \srII\ to include all three $t_{2g}$ orbitals as well as spin-orbit coupling~\cite{rag09,pue10,pue12}.

\begin{figure}[tb]
\includegraphics[width=7.5 cm]{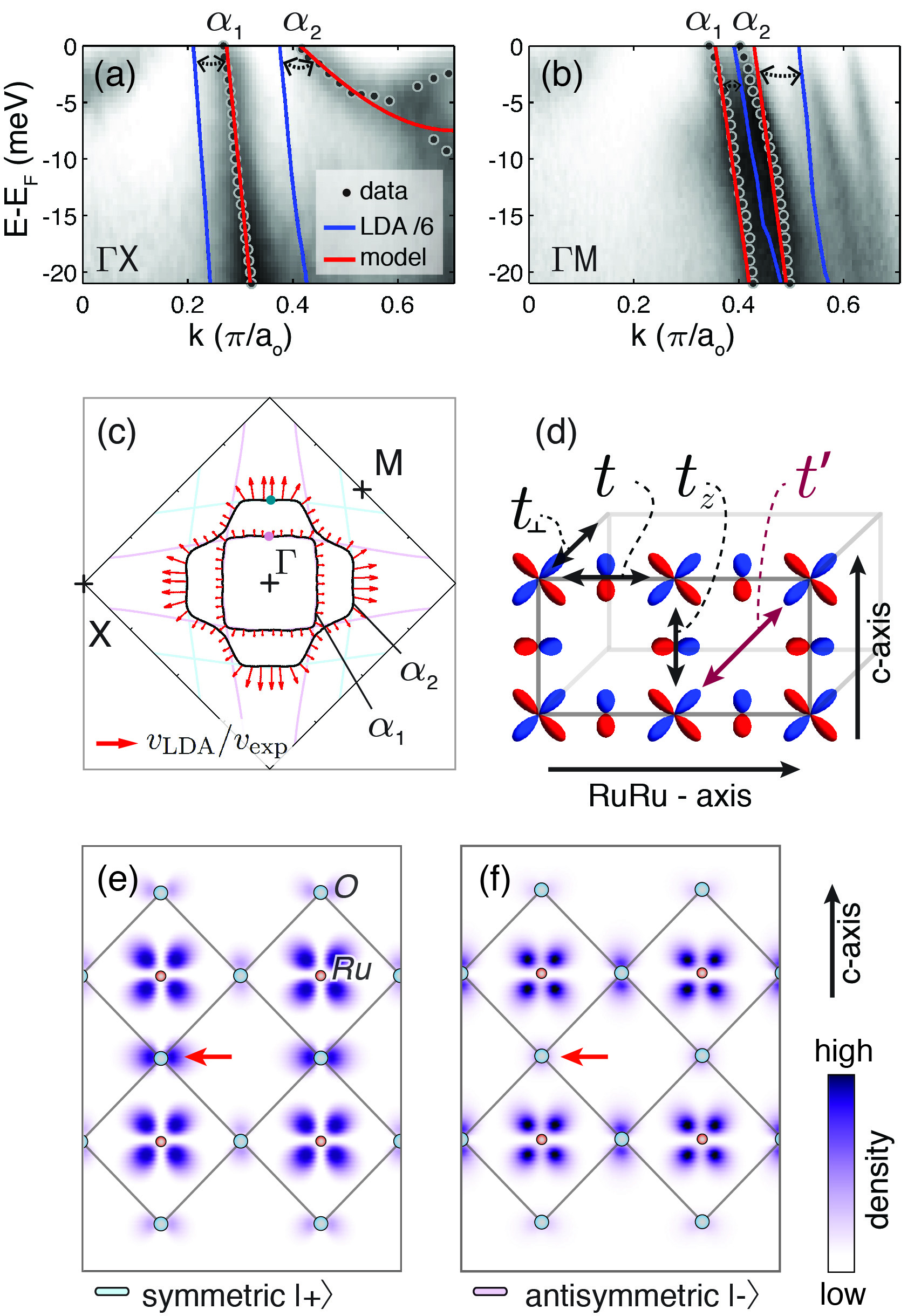}
\caption{Quasiparticle dispersion and renormalization along the $\alpha_{1,2}$ Fermi surface sheets. (a,b) show ARPES data along $\Gamma$X and $\Gamma$M with a globally compressed LDA calculation and a tight-binding dispersion overlaid. (c) Experimentally determined $\alpha_{1,2}$ Fermi surfaces. Red arrows are proportional to \zfl=$v_{\rm{LDA}}/v_{\rm{exp}}$, i.e. longer arrows indicate a higher renormalization constant. The bilayer-split fundamental bands from the one-dimensional \odxz, \odyz\ orbitals before hybridization are indicated in blue (symmetric $|+\rangle$) and pink (antisymmetric $|-\rangle$). (e,f) Valence charge density distribution for the highly and weakly renormalized states at $k$-points on the symmetric and antisymmetric Fermi surfaces (dots in (c)). The strongest difference is visible at the $O$ $p_z$ orbitals within a bilayer (red arrows).  }
\end{figure}
 
For the remainder of this paper we discuss the unusual Fermi surface sheet- and momentum-dependence of the mass enhancement in \srII\ which underlies the formation of the heavy $\alpha_2$ - $\gamma_2$ sheet. 
In Fig.~2(e) we compare the dispersion in \srII\ near the X$^{\rm{327}}$ point with the same $k$-space cut in \srI, which probes the heaviest single-layer states found in the $xy$ sheet around the unoccupied van Hove singularity at the M$^{\rm{214}}$-point. This shows that only the lightest bands in \srII\ are as dispersive as the bands in the single layer compound, confirming markedly enhanced correlations in \srII. However, the very large masses in \srII\ cannot be assigned exclusively to stronger interactions.
For instance in the case of the $xy$ sheets, the reduced dispersion of $xy'$ along $\Gamma$X$^{\rm{327}}$ is in part the simple consequence of back folding the isotropic fundamental $xy$ sheet resulting in Fermi surface contours that are nearly parallel to the $\Gamma$X$^{\rm{327}}$ line and thus disperse weakly along this direction.
This occurs in the LDA calculated bands as well, although their Fermi velocities are much higher. However, for the bilayer split quasi-one-dimensional $xz/yz$ and $xz'/yz'$ bands a different band width is unexpected and cannot be understood in a similar non-interacting band structure picture.

The $xz'/yz'$ bands hybridize weakly to form the $\alpha_{1,2}$ Fermi surface contours as shown in Fig.~3(c). Within LDA these sheets have nearly isotropic bare Fermi velocities, consistent with naive expectations. However, the quasiparticle dispersion along the $\alpha_2$ Fermi surface, which was also detected in spectroscopic imaging STM~\cite{lee09}, is clearly anisotropic. This is evident from the dispersion plots shown in Fig.~3(a,b), where we overlay an LDA dispersion, globally compressed by a factor of 6 onto the data. This reproduces the Fermi velocity along $\Gamma$M but overestimates it by a factor of $\approx 3$ along $\Gamma$X. The renormalization \zfl\ along the entire $\alpha_{1,2}$ Fermi surface sheets is visualized in panel (c) by red arrows with lengths proportional to $v_{\rm{LDA}}/v_{\rm{exp}}$, with $v_{\rm{exp}}$ extracted from a large number of cuts normal to the Fermi surface. This demonstrates that the self-energy in \srII\ has a marked momentum dependence, which is unexpected for a Fermi liquid where local electron-electron correlations dominate the mass enhancement. 

Intriguingly though, there is a clear correlation between \zfl\ and the character of the wave functions at the Fermi surface. To illustrate this we introduce a simple tight-binding model describing the $xz/yz$ orbitals of an isolated bilayer using in-plane $(t, t_{\perp})$ and out-of-plane $(t_z,t')$ hopping elements as illustrated in Fig.~3(d). The dispersion of this model along $\Gamma$X is $\varepsilon_{\bv{k}\pm} = 2(t\pm t')\cos k_{x} \pm t_{z}$ and its Fermi surface consists of the two sets of slightly curved light blue and pink lines in Fig.~3(c) which correspond to the symmetric ($|+\rangle$) and antisymmetric ($|-\rangle$) combinations of the $xz/yz$ orbitals in the upper and lower RuO$_2$ plane;  $|\pm\rangle= |upper\rangle \pm |lower\rangle$. 
It is evident from this analysis that the strongly and weakly renormalized bands exactly follow the $|+\rangle$ and $|-\rangle$ states, respectively. Since the antisymmetric combination $|-\rangle$ has a node at the apical O site, it is intuitive that these states hybridize to a different degree with O~$p$ levels.
Indeed, our LDA calculations (Fig.~3(e,f)) show that only the symmetric combination hybridizes strongly with apical O~$p_{x,y}$ states in between the two RuO$_2$ layers, whereas the antisymmetric combination has little weight at the apical O site but hybridizes with in-plane O~$p_z$ states. 
This has a moderate influence on the bare band width along $\Gamma$X which is around 1.6~eV for the antisymmetric and 0.9~eV for the symmetric combination corresponding to $t'/t\approx 0.25$ 
\footnote{The nearly identical LDA Fermi velocity of the $\alpha_{1,2}$ sheets along $\Gamma$X results from a compensation of the $k$-dependence of the group velocity along the band and the different band widths of the $|+>$, $|->$ states and is thus accidental.}
, similar to what is used for many theoretical models. 
Remarkably, our results suggest that interactions markedly amplify the difference in bare band width, possibly by reducing the effective in-plane hopping for the $|+\rangle$ states with little weight on in-plane oxygen. Within our empirical tight-binding picture, this corresponds to a strongly enhanced ratio $t'/t$ for the many-body dispersion of $\approx 0.85$. Thus, the complexity added over \srI\ by the RuO$_2$ bilayer in \srII\ goes far beyond simple band structure effects and includes substantially altered many-body interactions, which might hold the clue to their remarkably different thermodynamic properties.

We note that anisotropic renormalizations \zfl\ are often observed in single band systems where they are commonly attributed to coupling to bosonic modes. Neutron scattering detected strong antiferromagnetic spin fluctuations in \srII\ at energies $\lesssim$ 5~meV and wave vectors $|\bv{q}_1| =0.18$~$\pi/a_{\rm{t}}$ and $|\bv{q}_2| =0.5$~$\pi/a_{\rm{t}}$ \cite{cap03,ram08} that are a near perfect match to the nesting vectors connecting parallel sections of $\alpha_{1,2}$. 
However, we find a poor correlation between nesting and renormalization:
While $v_{\rm{LDA}}/v_{\rm{exp}}$ is clearly enhanced for the parallel sections of $\alpha_2$, this is not the case for the nearly square $\alpha_1$ sheet, which is as highly nested as $\alpha_2$ and whose nesting vector
matches the strongest antiferromagnetic fluctuations at $\bv{q}_2$ seen in neutron scattering. 
Moreover, both, the absolute magnitude of the renormalization in \srII\ as well as its large anisotropy are hard to reconcile with coupling to itinerant spin-fluctuations.

In conclusion, we illustrated how orbital dependent renormalization, backfolding and hybridization lead to the formation of heavy $d$-electron quasiparticles in \srII. 
We further argued that, in multi-band systems such as \srII, short range electron correlations can cause a pronounced anisotropy in the mass enhancement, which has a marked influence on the physical properties. We expect that both of these effects are of general relevance to 4$d$ transition metal oxides and other strongly correlated multi band systems.

We gratefully acknowledge discussions with 
A. Georges,
S. Raghu,
M.S. Golden,
R.G. Hennig,
C. Hooley,
J. Mravlje
A.W. Rost, 
S.C. Sundar,
and J. Zaanen.
This work has been supported by the Scottish Funding Council, the European Research Council and the UK EPSRC. SSRL is operated by the DOE's office of Basic Energy Science.

%

\end{document}